\documentclass[aps,prb,twocolumn,groupedaddress]{revtex4}
\usepackage{epsfig}
\usepackage{graphicx}
\usepackage{amsfonts}
\usepackage{amsmath,amssymb}
\usepackage{latexsym}
\usepackage{bm}

\newcommand{\beq}{\begin{eqnarray}}
\newcommand{\eeq}{\end{eqnarray}}

\begin{document}
\title{Monte Carlo studies of extensions of the Blume-Emery-Griffiths model}
\author{C.C. Loois, G.T. Barkema and C. Morais Smith}
\affiliation{Institute for Theoretical Physics, Utrecht
University, Leuvenlaan 4,3584 CE Utrecht, The Netherlands}
\date{\today}
\begin{abstract}
We extend the Blume-Emery-Griffiths (BEG) model to a two-component
BEG model in order to study 2D systems with two order parameters,
such as magnetic superconductors or two-component Bose-Einstein
condensates. The model is investigated using Monte Carlo
simulations, and the temperature-concentration phase diagram is
determined in the presence and absence of an external magnetic
field. This model exhibits a rich phase diagram, including a
second-order transition to a phase where superconductivity and
magnetism coexist. Results are compared with experiments on
Cerium-based heavy-fermion superconductors. To study cold atom
mixtures, we also simulate the BEG and two-component BEG models
with a trapping potential. In the BEG model with a trap, there is
no longer a first order transition to a true phase-separated
regime, but a crossover to a kind of phase-separated region. The
relation with imbalanced fermi-mixtures is discussed. We present
the phase diagram of the two-component BEG model with a trap,
which can describe boson-boson mixtures of cold atoms. Although
there are no experimental results yet for the latter, we hope that
our predictions could help to stimulate future experiments in this
direction.
\end{abstract}

\maketitle
\section{Introduction}

Mixtures of $^3$He and $^4$He atoms exhibit a rich phase diagram,
where besides a normal phase, there is a phase where $^4$He is
superfluid, and a phase separated region of superfluid $^4$He and
normal $^3$He.\cite{exphelium} In 1971, Blume, Emery and
Griffiths\cite{beg} proposed a model to describe such mixtures.
They simplified the continuous phase of the superfluid order
parameter such that it could acquire only two values. Although
they made this very rough approximation and modelled the uniform
system in a lattice, their results are very interesting.
Qualitatively, they reproduced the right phases and the right
orders of the phase transitions. Furthermore, if disorder is
introduced by placing the mixture into aerogel, after some
modifications,\cite{randomD1} the model can still yield the
experimentally observed phase diagram.\cite{randomD2}

Here, we generalize this model to a two-component case in order to
describe systems with two order parameters and study the problem
numerically, using Monte Carlo simulations. The motivation for the
model we are proposing is twofold. Firstly, we would like to study
condensed matter materials like heavy fermions, high-$T_c$
superconductors, and organic superconductors. In particular, we
want to study the interplay between magnetic and superconducting
ordering in these materials. Both order parameters are modelled as
an Ising spin variable. Concerning the magnetism, we consider the
ferro- and the antiferro-magnetic cases, and investigate also the
effect of an additional magnetic field. We find that in the
absence of a magnetic field, in the region where the two orders
coexist, the system is always phase separated. When we add a
magnetic field, we also find regions with microscopic coexistence
of the two phases. Secondly, we want to study mixtures of cold
atoms. Cold atoms have emerged in recent years as an ideal
simulator of condensed matter systems. Because experiments with
cold atoms are often carried out in a trap, we add a trapping
potential to the model. This fact qualitatively changes the
physics of the problem. For the case of a single component BEG
model in a trap, the results are compared with experimental and
theoretical work on imbalanced Fermi mixtures. For the case of the
two-component BEG model, we make predictions for the phase diagram
of boson-boson mixtures.

The outline of this paper is the following: in
section~\ref{sec:twobeg}, we introduce the two-component BEG
model, and investigate it in the presence and absence of an
external magnetic field. The effect of a trapping potential is
described in section~\ref{sec:trap}. In section~\ref{sec:comp}, we
compare the results with magnetic superconductors and cold atom
systems. Our conclusions are presented in section~\ref{sec:conc}.

\section{The two-component Blume-Emery-Griffiths model}
\label{sec:twobeg} The BEG model was originally proposed to
describe superfluidity.\cite{beg} The phase diagram found by Monte
Carlo simulations exhibits large similarities with the phase
diagram of $^3$He-$^4$He mixtures measured by
experimentalists.\cite{exphelium} The main idea of studying
superfluidity with the BEG model relies on the $U(1)$
symmetry-breaking of the ground-state wave function. For
superconductivity and Bose-Einstein condensation we have the same
symmetry breaking, hence we can try to model these phenomena in
the same way.

Several physical systems exhibit two unequal symmetry broken
phases simultaneously. A general Hamiltonian describing this class
of systems reads

\begin{equation}
{\cal H}=-J_1 \sum_{<ij>} \sigma_{i}\sigma_{j}-J_2 \sum_{<ij>}
s_{i}s_{j}+D\sum_{i} \sigma_{i}^{2} +H\sum_{i}\sigma_{i},
\label{eq:ham}
\end{equation}

\noindent where $(\sigma_i,s_i)$ can take the values
$(0,1),(0,-1),(1,0)$, and $(-1,0)$. This choice implies that only
one kind of boson can occupy each lattice site. $D$ is an
anisotropy field that controls the number of lattice sites with
nonzero $\sigma_i$. $H$ plays the role of an external magnetic
field, which may couple only to the order parameter describing a
magnetic transition. The Hamiltonian \eqref{eq:ham} is appropriate
for describing phase transitions which require two order
parameters, one describing the ordering of the fraction of the
system with nonzero $\sigma$, the other one of nonzero $s$. This
yields several possibilities, both fractions can model
superfluidity, superconductivity, or (anti)ferromagnetism.
Possible applications could be magnetic superconductors, or
two-component Bose-Einstein condensates.

From now on, we will consider the fraction with nonnegative
$\sigma$ as describing magnetism, and $s$ superconductivity
(preformed bosons that can Bose-Einstein condensate). Thus,
$\sigma_i$ represents the spin of particle $i$ and $s_i$ the
discretized phase of the wavefunction. Therefore, $J_1$ can be
both positive (ferromagnetism) and negative (antiferromagnetism),
but $J_2$ has to be positive. We define the concentration, the
ferromagnetic, antiferromagnetic, and superconducting order
parameters as
\begin{eqnarray}
c=\frac{1}{N}\sum_{i} \sigma_{i}^{2},\\
m_{\rm{fm,\sigma}}= \frac{1}{N}\sum_{i}\sigma_{i},\\
m_{\rm{af,\sigma}}=\frac{1}{N}\sum_{i}(-1)^i\sigma_{i},\\
m_s=\frac{1}{N}\sum_{i} s_{i}.
\end{eqnarray}
Note that $m_{\rm{fm,\sigma}}$ and $m_{\rm{af,\sigma}}$ can reach
a maximum value of $c$, and $m_s$ of $1-c$. We define the ratio
between the two coupling constants $J_2$ and $J_1$ as
\begin{equation}
K=\frac{J_2}{|J_1|}.
\end{equation}

\subsection{The Method}
We investigate this model by Monte Carlo simulations. To determine
the location of second-order phase transitions, we performed
simulations at constant concentration, in which the elementary
moves were flips of $s_i$ and $\sigma_i$ or nonlocal spin
exchanges. The location of the transition is then obtained from
the peak location of the magnetic susceptibility. The locations of
first-order phase transitions are obtained from simulations at
constant temperature, with as elementary moves local flips of
$s_i$ and $\sigma_i$, as well as same-site replacements of $s_i$
by $\sigma_i$ and vice versa. A jump in the concentration $c$ as a
function of the anisotropy field $D$ is then the signature of the
phase transition.

All simulations are performed on lattices with approximately
$40\times 40$ sites. Per point in the phase diagram, simulations
were run over $3\cdot 10^5$ to $3\cdot 10^7$ Monte Carlo steps per
site, depending on the correlation times.

\subsection{Zero magnetic field, $H=0$} In the absence of a
magnetic field, the Hamiltonian \eqref{eq:ham} has
ferro-antiferromagnetic symmetry.

\begin{figure}[!htb]
\centerline{
\mbox{\includegraphics[width=0.4\textwidth,angle=0]{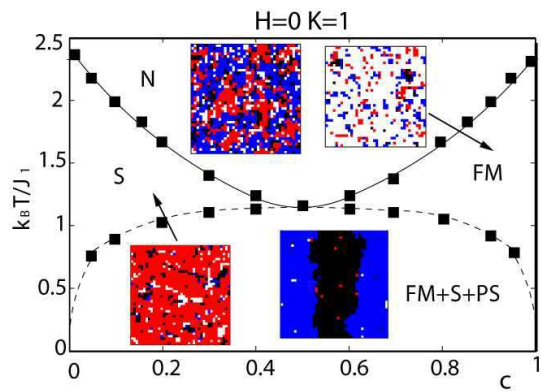}}
}\caption{(color online) Phase diagram, temperature (in units of
$J_1/k_B$) versus concentration, in the absence of a magnetic
field. N indicates the normal phase, S superconductivity, FM
ferromagnetism, and PS phase separation. Solid lines represent
second order phase transitions, dashed lines first order ones.
Lines are guides to the eye. Snapshots of the simulation are
shown. Black (white) represents $\sigma_{i}=1$ $(-1)$, red (blue)
$s_i=1$ $(-1)$. }\label{figure:H0K1}
\end{figure}

First, we consider $K=1$. In this case, $J_1=J_2$, and the shape
of the phase diagram must be symmetric under the transformation $c
\to 1-c$. The results of the simulations are plotted in
Fig.~\ref{figure:H0K1}. We see that it indeed obeys this symmetry
and exhibits four phases: a superconducting phase (S), where the
order parameter $m_s$ is nonzero, a ferromagnetic phase (FM),
where $m_{\rm{fm,\sigma}}$ is nonzero, a phase-separated regime
(PS) where the spins and the angular phases have formed ordered
clusters, and finally the normal phase (N), in which there is
neither order nor phase separation. Analogous to the BEG model,
the transition from the phase-separated regions to other phases
are first-order (dashed line), the other ones are second-order
(continuous line).

\begin{figure}[!htb]
\centerline{
\mbox{\includegraphics[width=0.4\textwidth,angle=0]{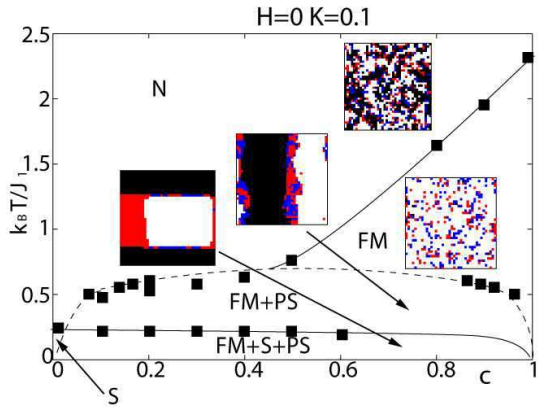}}
}\caption{(color online) Phase diagram in the absence of a
magnetic field, for a relative coupling constant of $K=0.1$.
}\label{figure:H0K01}
\end{figure}

Second, we consider the case $K=0.1$. The results of the
simulations are plotted in Fig.~\ref{figure:H0K01}. We can
understand the results as follows: $J_2$ is much smaller than
$J_1$, hence the spins will not pay much attention to the angular
phases, and the part of the phase diagram concerning the spins
will be very similar to the BEG model. Because $J_2$ is so small,
the phases will only order at very low temperatures (at zero
concentration, the temperature is ten times lower than the one at
which the spins order at a concentration of one). If the
concentration is slightly raised from zero, the system is already
in the phase separated regime. All the states with a nonzero phase
have clustered, and are not diluted by states with nonzero spin.
Therefore, the critical temperature in the phase separated region
will approximately remain constant. Because the temperature at
which the angular phases order is lower than the temperature at
which phase separation begins, there is a phase separated region
in which the angular phases of the wavefunction are not ordered,
which may appear unexpected at first sight. The transition within
the phase separated regime, from the region where the angular
phases are not ordered to the phase where they are ordered
(superconductivity), is second-order. This is expected, because in
the phase separated regime, all the phases have clustered, and the
transition will be comparable with the transition in the Ising
model, which is also second-order.
\subsection{Adding a magnetic field: the
antiferromagnetic case}
 If we apply a nonnegative uniform magnetic
field to the system, the ferro-antiferromagnetic symmetry is
broken. We choose to consider the antiferromagnetic case here,
because then there are two competing effects, the magnetic field
tends to align the spins, whereas the exchange interaction wants
to order the spins antiferromagnetically. The magnetic field $H$
will be measured in units of $J_1$.

Kimel {\it{et al.}}\cite{dh3} have studied the antiferromagnetic
BEG model in the presence of a magnetic field, using Monte Carlo
simulations. Their results at zero temperature suggest that the
behavior of the system should be separated into three
qualitatively distinct regions, namely $H \in [0,2], H \in[2,4]$
and $H \in[4,\infty]$. We consider here the cases $K=1$ and
$K=0.1$ for values of $H$ within each of these intervals.

\subsubsection{H=1.5} First, we considered a magnetic field in the
interval $[0,2]$, namely $H=1.5$. Both for $K=1$ and $K=0.1$, the
results (not shown) are qualitatively the same as in the case of
$H=0$. This behavior was expected from the phase diagram of the
single-component BEG model at zero temperature. Because the
magnetic field tries to align the spins, the antiferromagnetic
transition temperature is lower than in the absence of a magnetic
field.

\subsubsection{H=2.5}

In the usual BEG model, the first-order phase transition
disappears in the presence of a magnetic field $H \in [2,4]$. At
zero temperature, there is a second-order phase transition between
a state with $\sigma_i=0$ at every site, and a checkerboard phase,
where one sublattice has $\sigma_i=0$ at every site, and the other
one $\sigma_i=-1$. There is also a transition between the
checkerboard state, and an antiferromagnetic phase, but this
transition is absent at nonzero temperature.\cite{dh3}

\begin{figure}[!htb]
\centerline{
\mbox{\includegraphics[width=0.4\textwidth,angle=0]{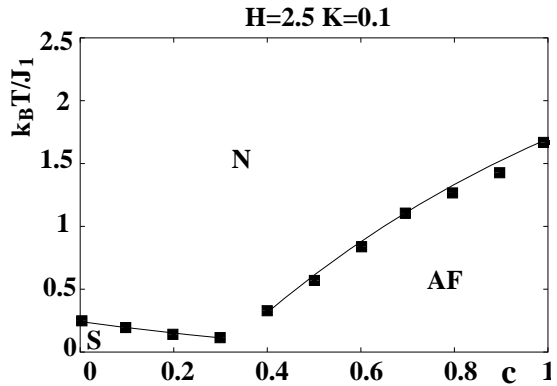}}
}\caption{Phase diagram at a magnetic field $H=2.5$ and relative
exchange strength $K=0.1$. N denotes the normal phase, S
superconductivity and AF antiferromagnetism.
}\label{figure:H25K01}
\end{figure}

\begin{figure}[!htb]
\centerline{
\mbox{\includegraphics[width=0.4\textwidth,angle=0]{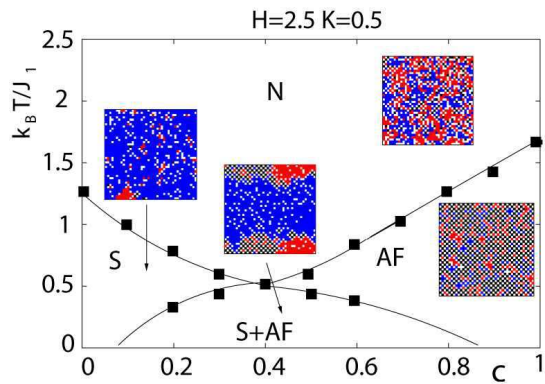}}
}\caption{(color online) Phase diagram at a magnetic field $H=2.5$
and relative exchange strength $K=0.5$. There is a region where
superconductivity and antiferromagnetism coexist, but where there
is no true phase separation. }\label{figure:H25K05}
\end{figure}

For $K=1$, the behavior of the two-component BEG model is still
very similar to the case $H=0$. For $K=0.1$, the first-order phase
transition disappears, and therefore there is no phase-separated
region, see Fig.~\ref{figure:H25K01}. We do observe an
antiferromagnetic and a superconducting phase, but it is not clear
from the figure whether the two phases overlap. To better
understand this low-$T$ intermediate regime, we also simulated the
problem at a relative coupling strength of $K=0.5$. In
Fig.~\ref{figure:H25K05}, we clearly observe that there is a
region where antiferromagnetism and superconductivity coexist,
{\it without true phase separation}, since the first-order phase
transition has disappeared. What is also interesting is that at
zero temperature this region begins at a nonzero concentration,
and ends at a concentration smaller than one. When there is phase
separation, this coexistence region always begins at $c=0$ and
ends at $c=1$.

\subsubsection{H=5} 
\begin{figure}[!htb]
\centerline{
\mbox{\includegraphics[width=0.4\textwidth,angle=0]{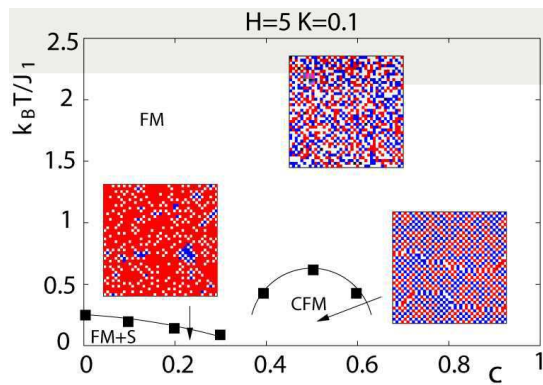}}
}\caption{(color online) Phase diagram at $H=5$ and $K=0.1$. FM
denotes ferromagnetism, S superconductivity and CFM denotes
checkerboard ferromagnetism. }\label{figure:H5K01}
\end{figure}

In the original BEG model, when the magnetic field is increased
to a value higher than $H=4$ at zero temperature,
antiferromagnetism totally disappears because the spins tend to
align with the magnetic field.\cite{dh3} The system is therefore
magnetized, but not because of the nearest-neighbor interactions.
Therefore, this is not really ferromagnetism, but for the sake of
simplicity, we denote it like this. For the case of $K=1$, we
observe a phase with ferromagnetic and superconducting ordering,
and a ferromagnetic phase (not shown). For $K=0.1$, we find
another interesting phase, namely a ferromagnetic checkerboard
phase, consisting of two sublattices, see Fig.~\ref{figure:H5K01}.
At the first sublattice, all sites are randomly occupied by phases
with a value of $s_i=1$ or $s_i=-1$. At the second one, all sites
are occupied by the spin that is favored by the magnetic field,
$\sigma_i=-1$. This phase is most likely to occur at a
concentration of $c=0.5$ because in this case a perfect
checkerboard is possible.
\section{Adding a trap potential}
\label{sec:trap}
\subsection{The Blume-Emery-Griffiths model}
\label{sec:trapbc} Because experiments with cold atoms are often
carried out in a trap, we will add a harmonic potential to the
original BEG Hamiltonian, to describe mixtures of fermions and
bosons in a trap. In general, the potential felt by the bosons is
different from the one felt by the fermions, what implies that we
must include two terms,
\begin{equation}
a_b\sum_i (x_i^2+y_i^2)\sigma_i^2 + a_f\sum_i
(x_i^2+y_i^2)(1-\sigma_i^2).
\end{equation}
 Here, $x_i$ and $y_i$ are the horizontal and vertical
 distances of site $i$, measured from the center of the lattice, in lattice
 units, and $a_b$ and $a_f$ measure how much the bosons (the
 states with $\sigma_i=\pm1$), and the fermions (the states with
 $\sigma_i=0$) feel the influence of the trap. If $a_b=a_f$, this
 term is constant, and the phase diagram is not modified. We will consider the case
 $a_b>a_f$, which is the most relevant experimentally. Using the hard core constraint
 $\sigma_i^2+s_i^2=1$, we can then rewrite this term and add it to the
 BEG Hamiltonian, thus obtaining
\begin{equation}
{\cal H}=-J\sum_{<ij>} \sigma_{i}\sigma_{j} + D\sum_{i}
\sigma_{i}^{2}+a \sum_i (x_i^2+y_i^2)\sigma_i^2,
\end{equation}

\noindent where $a=a_b-a_f$. This means that, effectively, the
bosons will feel a stronger tendency to go to the center of the
trap.

\begin{figure}[!htb]
\centerline{
\mbox{\includegraphics[width=0.4\textwidth,angle=0]{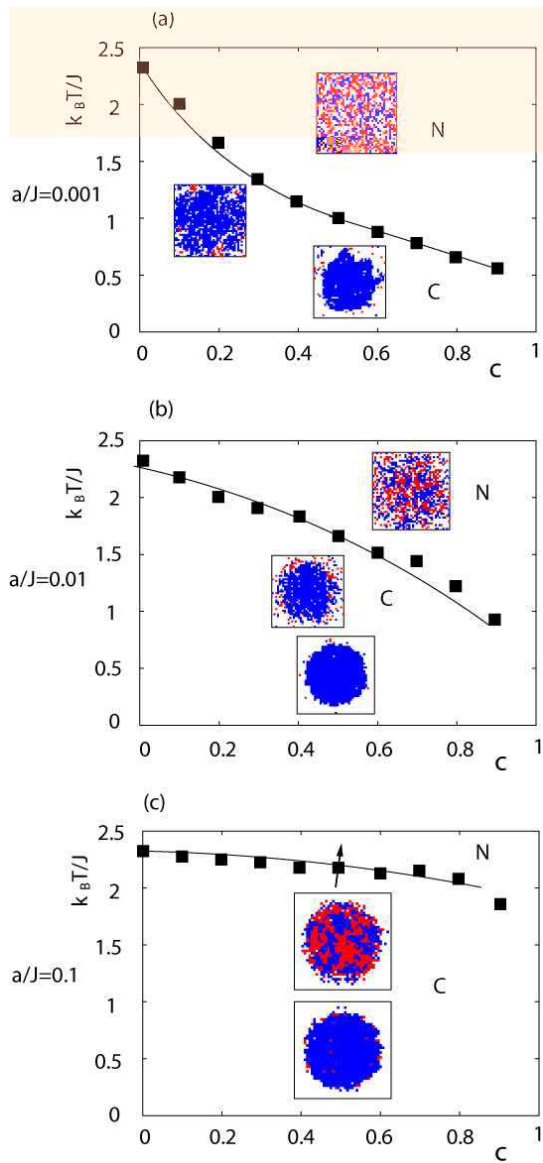}}
}\caption{(color online) Phase diagrams of the BEG model with a
trapping potential. $N$ denotes the normal, unordered state, $C$
the condensed phase, in which the sites with $\sigma_i=\pm 1$ are
ordered. Lines are guides to the eyes. Snapshots are shown, where
blue represents $\sigma_i=1$, red $\sigma_i=-1$, and white
$\sigma_i=0$.}\label{figure:trap1beg}
\end{figure}

In the limit of $a\to \infty$,
 all the states with $\sigma_i=\pm 1$ will cluster in the center
 of the trap, and therefore the ordering temperature will be the
 same as in the Ising model. Note that the maximum value of the extra
 term in the Hamiltonian  will depend on the size of the lattice. This
 way
 of including the trapping potential is comparable with the work
 of Gygi {\it et al.},\cite{Gygi} where a spatial-dependent chemical potential was added to
 the Bose-Hubbard model in order to describe bosonic atoms in an optical
 lattice.

We simulated the new model using the same procedures as for the
BEG model and the two-component BEG model. The results for three
different strengths of the trapping potential are plotted in
Fig.~\ref{figure:trap1beg}. In the BEG model without a trap, there
is a second-order phase transition from a normal state to an
ordered state, and a first-order phase transition to a
phase-separated region\cite{beg}. For the three values of $a$
considered here, we do not find a first-order phase transition any
more. A part of the first-order phase transition line disappears,
and a part changes into a second-order one.

We see that for a small difference between the trap potential felt
by the bosons and the fermions, $a/J=0.001$, the transition
temperatures are very similar to the case without a trap. For a
large difference, $a/J=0.1$, the transition temperatures approach
the transition temperature of the Ising model for almost all
concentrations, as expected. When the states with $\sigma_i=\pm 1$
are ordered, we will not speak of a superfluid state, but of a
condensed state, because we now consider bosons in general.

It is important to estimate at which temperature the system starts
to feel the influence of the trapping potential. Let us assume
that a cluster of size $m$ feels the potential when the energy
difference between the state with this cluster in the center and
in the corner of the lattice is of the order $k_BT/J$. For a
lattice of size $L^2$, this estimation results in
\begin{equation}
\label{eq:est} \frac{maL^2}{2J}\sim \frac{k_BT}{J}.
\end{equation}
In this approximation, a single particle ($m=1$) in a lattice of
size $L=41$ will start to feel the potential if $k_BT/J\sim 800a$.
For $a/J=0.1$ and $a/J=0.01$ this results in $k_BT/J\sim 80$ and
$k_BT/J\sim 8$, respectively, in both cases much higher than the
temperatures we are interested in, because ordering starts around
$k_BT/J\approx 2.4$. Therefore, the single particles will
experience the influence of the trap in the entire temperature
range of Fig.~\ref{figure:trap1beg} $(b)$ and $(c)$. For
$a/J=0.001$, a single particle will feel the trapping potential
for temperatures lower than $k_BT/J \sim 0.8$. However, for higher
temperatures the system already orders, and therefore there are
some large clusters that according to Eq. (\ref{eq:est}) will feel
the potential already at much higher temperatures. This reasoning
is in agreement with the snapshots in Fig.~\ref{figure:trap1beg}
$(a)$. For $a/J=0.001$, we clearly observe the influence of the
trap when the states with $\sigma_i=\pm 1$ have clustered. In the
disordered state, the influence is less visible. For $a/J=0.1$ and
$a/J=0.01$, we indeed see the influence of the trap for all
temperatures, even in the disordered state.

\subsection{The two-component Blume-Emery-Griffiths model}
Analogous to the previous subsection, we will also add a trapping
potential to the two-component BEG model. In the latter, both the
states with $\sigma_i=\pm 1$ and $s_i=\pm 1$ describe bosons, that
both can condense. Therefore, this model can be applied to study
cold atoms mixtures with two species of bosons. We will consider
the realistic case that the two species feel different trapping
potentials. Therefore, we add the extra terms
\begin{equation}
a_{\sigma} \sum_i (x_i^2+y_i^2)\sigma_i^2+a_{s} \sum_i
(x_i^2+y_i^2)s_i^2
\end{equation}
to the Hamiltonian. Because at every lattice site
$\sigma_i^2+s_i^2=1$, we can rewrite this term and add it to the
two-component BEG Hamiltonian, to get

\begin{eqnarray}
{\cal H}&=&-J_1 \sum_{<ij>} \sigma_{i}\sigma_{j}-J_2 \sum_{<ij>}
s_{i}s_{j}+D\sum_{i} \sigma_{i}^{2}\nonumber\\ && +a \sum_i
(x_i^2+y_i^2)\sigma_i^2,
\end{eqnarray}
where $a=a_{\sigma}-a_s$ is now the difference between the
potentials felt by the two species of bosons. Now, the bosons with
$\sigma_i=\pm 1$ have a stronger tendency to go to the center of
the lattice.

\begin{figure}[!htb]
\centerline{
\mbox{\includegraphics[width=0.4\textwidth,angle=0]{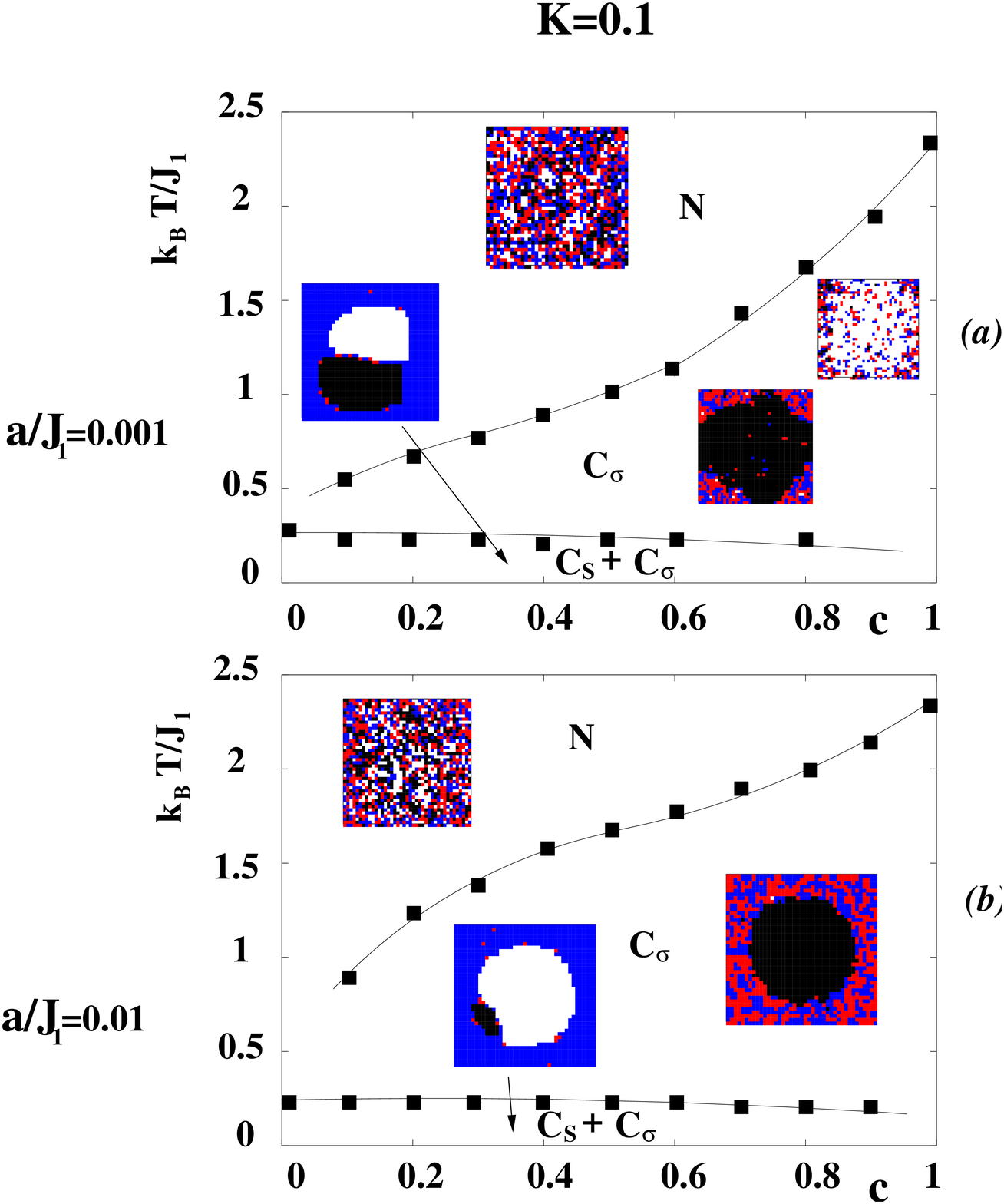}}
}\caption{(color online) Phase diagrams of the two-component BEG
model with a trapping potential. $N$ denotes the normal, unordered
state, $C_{\sigma}$ and $C_s$ the phases where the bosons
represented by the state with $\sigma_i=\pm 1$, respectively
$s_i=\pm 1$ are condensed. Lines are guides to the eyes. Snapshots
are shown, where black and white represent $\sigma_i=\pm 1$, and
red and blue $s_i=\pm 1$.}\label{figure:trapbeg2K01}
\end{figure}

\begin{figure}[!htb]
\centerline{
\mbox{\includegraphics[width=0.4\textwidth,angle=0]{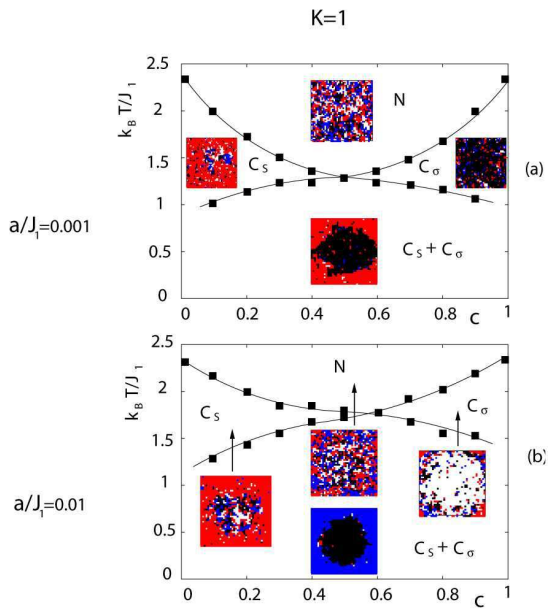}}
}\caption{(color online) Phase diagrams of the two-component BEG
model with a trapping potential. The notation used is the same as
in Fig.~\ref{figure:trapbeg2K01}.}\label{figure:trapbeg2K1}
\end{figure}
The results of our simulations are plotted in
Figs.~\ref{figure:trapbeg2K01} and~\ref{figure:trapbeg2K1}. We
considered two different strengths of the trapping potential, and
two different ratios of the coupling strengths of the bosons,
namely $K=J_2/J_1=1$ and $K=0.1$. For $K=0.1$, the right part of
the first-order phase transition disappears and the left one
becomes second-order (see Fig.~\ref{figure:trapbeg2K01}), whereas
for $K=1$ both left and right
parts of the first-order phase transition are converted into second-order (see Fig.~\ref{figure:trapbeg2K1}).\\
In the limit of $a\to \infty$, all the sites with $\sigma_i=\pm 1$
will have clustered in the center of the lattice, and all sites
with $s_i=\pm 1$ at the corners. Therefore, for all
concentrations, the system behaves as two uncoupled Ising models.
In the case of $K=1$, we see indeed that the transition
temperatures for both species approach the Ising transition
temperature. For $K=0.1$, because $J_2$ is ten times smaller than
$J_1$, one of the species will order at the Ising transition
temperature, and the other one at one tenth of the Ising
transition temperature.

To find the temperature at which the system starts to feel the
presence of the trap, we can make the same analysis as in
subsection~\ref{sec:trapbc}. Also here, we see in the snapshots of
Figs.~\ref{figure:trapbeg2K01} and~\ref{figure:trapbeg2K1} that
for $a/J_1=0.1$ (not shown) and $a/J_1=0.01$, the system always
feels the influence of the trap, and for $a/J_1=0.001$, it does
only when the system is ordered.
 If we inspect Fig.~\ref{figure:trapbeg2K1}
$(a)$, we see that there is a phase $C_s$ in which the bosons
represented by $s_i=\pm 1$ are ordered, but the bosons represented
by $\sigma_i=\pm 1$ are not. This is somewhat surprising. A reason
for the occurrence of this phase is that when all the bosons that
have the tendency to go to the center of the trap have clustered
there, automatically also the other bosons have clustered at the
edge. Therefore, they can have nearest-neighbor interactions, and
they can easily order. It remains to see whether such a phase
indeed occurs in experiments. From the theoretical point of view,
it would be interesting to also allow for states with
$\sigma_i=s_i=0$, to verify the stability of this phase, when we
relax the constraint that every lattice site must be occupied by
one of the bosons.  Note that for small enough concentrations,
this phase will always occur, since the bosons $s$ are hardly
diluted by the bosons $\sigma$.
\section{Comparison with experiments}
\label{sec:comp}
\subsection{Magnetic superconductors}

\begin{figure}[!htb]
\centerline{
\mbox{\includegraphics[width=0.4\textwidth,angle=0]{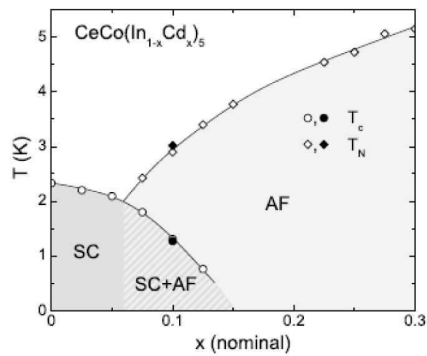}}
}\caption{Phase diagram of CeCo(In$_{1-x}$Cd$_x$)$_5$. The figure
is extracted from Ref.~\onlinecite{nick1}. }\label{figure:ceco}
\end{figure}
There are several examples of Cerium-based superconductors, for
example CeCoIn$_5$ and CeIrIn$_5$, as well as antiferromagnets
that contain this element, like CeRhIn$_5$, CeCoCd$_5$,
CeRhCd$_5$, and CeIrCd$_5$. Let us consider CeCoIn$_5$ and
CeCoCd$_5$. These two materials have two elements in common, Ce
and Co, and differ in the third element. By doping CeCoIn$_5$ with
Cd on the In site, we can change the superconductor CeCoIn$_5$
into an antiferromagnet. There are more of these Cerium-based
pairs, and therefore, this class of materials is appropriate for
studying the interplay between superconductivity and magnetism.

Let us consider the heavy fermion superconductor CeCoIn$_5$, with
Cadmium doping on the In-site. This material has the highest
superconducting transition temperature ($T_c=2.3K$) of all heavy
fermions, and its electronic structure is quasi-2D.\cite{quasi}
Nicklas {\it et al.}\cite{nick1} and Pham {\it et
al.}\cite{reversible} determined the antiferromagnetic and
superconducting onset temperatures of this material as a function
of doping by elastic neutron scattering, specific heat, and
resistivity measurements. Their results are plotted in
Fig.~\ref{figure:ceco}. For experimental details we refer the
reader to Ref.~\onlinecite{nick1}. The phase diagram of
CeCo(In$_{1-x}$Cd$_x$)$_5$ shows three ordered phases: a
superconducting phase, a commensurate antiferromagnetic phase, and
a region where superconductivity and antiferromagnetism
microscopically coexist.

\begin{figure}[!htb]
\centerline{
\mbox{\includegraphics[width=0.4\textwidth,angle=0]{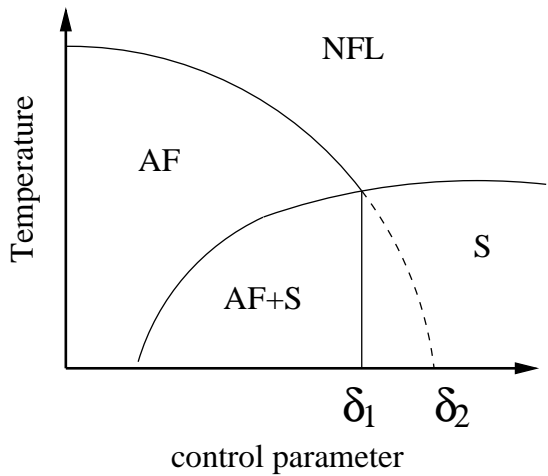}}
}\caption{Schematic phase diagram of unconventional
superconductors in temperature-control parameter space. AF denotes
antiferromagnetism, S superconductivity and NFL a non-Fermi
liquid. Experimentally, antiferromagnetism often disappears
abruptly at some critical value $\delta_1$ of the control
parameter, although one would expect a magnetic quantum critical
point at some value $\delta_2$ of the control parameter.
}\label{figure:artikel3}
\end{figure}

It is interesting to observe that in this material
antiferromagnetism suddenly disappears at the point where the
onset temperatures for superconductivity and antiferromagnetism
are equal. This feature, however, may change in the presence of an
applied magnetic field. In Fig.~\ref{figure:artikel3} we see a
schematic phase diagram of unconventional superconductors, in
temperature-control parameter space. In the case of
CeCo(In$_{1-x}$Cd$_x$)$_5$, the control parameter would be doping.
Another example of such a parameter is pressure. Park {\it et
al.}\cite{park} determined the phase diagram of CeRhIn$_5$ in
temperature-pressure space with and without a magnetic field.
Without a magnetic field, they also found this abrupt
disappearance of the incommensurate antiferromagnetic order at
$\delta_1$. However, when they applied a field of 33 KOe, the line
of the magnetic ordering temperature went smoothly down to zero at
$\delta_2$. Such a phase diagram shows many similarities with
Fig.~\ref{figure:H25K05} if we identify pressure with inverse
concentration in our model. Indeed, for an external magnetic field
of $H=2.5$ and a relative exchange strength $K=0.5$ (see
Fig.~\ref{figure:H25K05}), the phase diagram shows the same three
ordered phases. Further, the coexisting phase is not phase
separated.

\begin{figure}[!htb]
\centerline{
\mbox{\includegraphics[width=0.4\textwidth,angle=0]{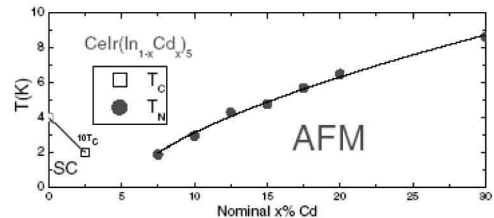}}
}\caption{Phase diagram of the heavy fermion
CeIr(In$_{1-x}$Cd$_x$)$_5$. The figure is extracted from
Ref.~\onlinecite{reversible}.}\label{figure:ceir}
\end{figure}

Finally, we consider the compound CeIr(In$_{1-x}$Cd$_x$)$_5$, see
Fig.~\ref{figure:ceir} and Ref.~\onlinecite{reversible}. For this
material, it is not clear if there is a region where
superconductivity and magnetism coexist. If there is such a
region, it is in a small doping interval. The phase diagram of
this material strongly resembles the phase diagram of the
two-component BEG model with an external magnetic field of $H=2.5$
and a relative coupling strength of $K=0.1$, see
Fig.~\ref{figure:H25K01}. Although this experiment was also
carried out without an external magnetic field, we only find
similarities with our model in the presence of a magnetic field.

\subsection{Cold atom systems}

\begin{figure}[!htb]
\centerline{
\mbox{\includegraphics[width=0.4\textwidth,angle=0]{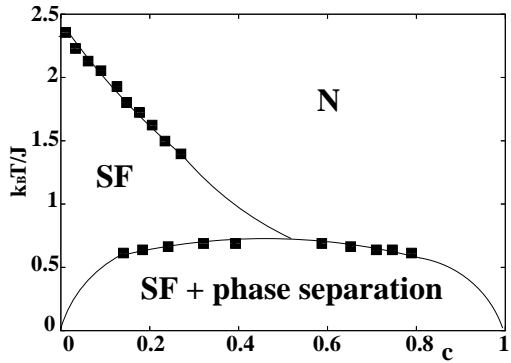}}
}\caption{Phase diagram of the original BEG model, obtained by
Monte Carlo simulations. $N$ denotes the normal phase, $SF$
superfluidity. Lines are guides to the eyes. The transition
between the normal and the superfluid state is second-order, the
transition to the phase separated regime is first-order.
}\label{figure:beg}
\end{figure}

In 2006, two experimental groups, the group of Ketterle at
MIT,\cite{MIT} and the group of Hulet at Rice
University,\cite{Rice} have performed experiments with imbalanced
ultracold $^6$Li atoms in a trap, and obtained contradictory
results. The MIT group measured a transition between a normal and
a superfluid phase at a polarization of $P\approx 0.70$, whereas
the group at Rice University observed a transition between two
superfluid phases at $P\approx 0.09$. Here, $P$ measures the
imbalance between the spin-up and the spin-down atoms,
\begin{equation}
P=\frac{N_{\uparrow}-N_{\downarrow}}{N_{\uparrow}+N_{\downarrow}}.
\end{equation}
Gubbels {\it et al.}\cite{koos} have set up a theoretical model to
describe these imbalanced Fermi mixtures and determined a general
phase diagram in temperature-polarization space that can explain
the observations of both groups. The topology of their phase
diagram shows large similarities with the phase diagram of the BEG
model. We can understand this resemblance as follows. In the BEG
model, the concentration $c$ is the fraction of lattice sites with
$\sigma_i=0$, and thus the fraction of the system that cannot
condense. The polarization $P$ is a measure for the difference of
the atoms in the spin-up and the spin-down state, and thus for the
number of fermions that remain after the others have paired. The
atoms with spin up and spin down will form pairs, and such a pair
can be described as a boson. Therefore, the polarization is also a
measure for the fraction of the system that cannot condense, and
the concentration can be mapped onto the polarization. We can
identify the paired atoms, the preformed bosons, with the states
$\sigma_i=\pm1$, and the remaining fermions with $\sigma_i=0$, see
Fig.~\ref{figure:beg}.

The experiments with $^6$Li are carried out in a trap, and the
theoretical model of Gubbels {\it et al.} only includes the
presence of the trap by using the local density approximation.
Now, we would like to compare their phase diagram with our results
of the BEG model with a trapping potential, in
Figs.~\ref{figure:trapbeg2K01} and~\ref{figure:trapbeg2K1}.
Although in the case of imbalanced fermions the frequency $\omega$
of the optical trap felt by the pairs of fermions (bosons) and the
remaining unpaired fermions is the same, the mass of the bosons is
twice as large, and the potential constant $a_b$ is thus larger
than $a_f$. This means that the comparison must be made with the
BEG model in a trap. In this model, the first-order phase
transition, measured by a jump in the concentration as a function
of the anisotropy field $D$ has disappeared, thus there is no true
transition to a phase-separated regime. However, if we inspect the
snapshots, we see that for low enough temperatures, or large
enough trapping potential, there still is a clear separation
between the condensed bosons and the fermions, suggesting some
kind of effective phase separation. We note that in experiments,
phase separation is measured by inspecting the radii of the clouds
of the atoms in the different hyperfine states, and not by a jump
in some order parameter.\cite{Rice} Our results thus suggest that
the measured different radii are not per se an evidence of a true
thermodynamic phase separation. Further experiments are required
to clarify this issue.

Although our model describes qualitatively the experimentally
observed phases, it cannot capture the fine details of recent
experimental results. Studies by Shin {\it et al.}\cite{ketter}
indicate that there is no superfluid phase, or phase-separated
phase for polarizations above $P\approx 0.36$. By a quantum Monte
Carlo approach, Lobo {\it et al.}\cite{lobo} predict a phase
transition between a normal and a superfluid state at a
polarization of $P\approx 0.39$ at zero temperature, and Gubbels
and Stoof\cite{koos2} recovered this results using a Wilsonian
renormalization group theory.

\section{Conclusions}

\label{sec:conc} We simulated a two-component extension of the BEG
model without an external magnetic field and determined the phase
diagram in the concentration-temperature space. In the region
where magnetism and superconductivity coexist, the system is
always phase separated. We added a magnetic field to our model,
and considered the antiferromagnetic case. In this case, we also
find phase diagrams with true coexistence of two ordered phases.
These diagrams are comparable with the phase diagram of doped
heavy fermions in the presence of a magnetic field.

In order to describe cold atom systems, we added a trapping
potential to the BEG model, and our extension of this model. The
added potential changes the phase separation regime conceptually.
We cannot speak anymore about true phase separation, but more
about a crossover to a phase separated region. We argue that the
BEG model with a trapping potential can be used to model
imbalanced Fermi mixtures. However, there are still quantitative
differences with experiments, which our model is not able to
cover. We also made predictions for the phase diagram of
boson-boson mixtures based on our simulations of the two-component
BEG model with a trapping potential. Although there is no
available experimental data on boson-boson mixtures, we hope that
our work can motivate further studies in this direction.

\section{Acknowledgements}
We are grateful to K. Gubbels, H. van Beijeren, A. de Vries and R.
Movshovich for fruitful discussions and to J. de Graaf for
technical help.

\end{document}